# Annotation as a New Paradigm in Research Archiving

Two Case Studies: Republic of Letters - Hebrew Text Database

**Dirk Roorda**
Data Archiving Networked Services (KNAW)
P.O. Box 93067
2509 AB Den Haag
Netherlands
dirk.roorda@dans.knaw.nl

**Charles van den Heuvel**
Huygens ING (KNAW)
P.O. Box 90754
2509 LT Den Haag
Netherlands
charles.van.den.heuvel@huygens.knaw.nl

KNAW = Royal Netherlands Academy of Arts and Sciences, knaw.nl

**ABSTRACT**

We outline a paradigm to preserve results of digital scholarship, whether they are query results, feature values, or topic assignments. This paradigm is characterized by using annotations as multifunctional carriers and making them portable. The testing grounds we have chosen are two significant enterprises, one in the history of science, and one in Hebrew scholarship.

The first one (CKCC) focuses on the results of a project where a Dutch consortium of universities, research institutes, and cultural heritage institutions experimented for 4 years with language techniques and topic modeling methods with the aim to analyze the emergence of scholarly debates. The data: a complex set of about 20.000 letters.

The second one (DTHB) is a multi-year effort to express the linguistic features of the Hebrew bible in a text database, which is still growing in detail and sophistication. Versions of this database are packaged in commercial bible study software.

We state that the results of these forms of scholarship require new knowledge management and archive practices. Only when researchers can build efficiently on each other's (intermediate) results, they can achieve the aggregations of quality data by which new questions can be answered, and hidden patterns visualized. Archives are required to find a balance between preserving authoritative versions of sources and supporting collaborative efforts in digital scholarship. Annotations are promising vehicles for preserving and reusing research results.

**INTRODUCTION**

In the Early Modern History of Europe, letters were by far the most important means of communication and played a role in the emergence of scholarly communities. Although from the 1660s onwards this one-to-one means of exchange of knowledge was gradually replaced by a more public form of scholarly communication via learned periodicals, communication via letters did continue. Given the role of the letter in scholarly communication and the emergence of scientific communities in Europe it is not surprising that the so-called "Republic of Letters" became a recurrent theme in the history of the humanities and sciences.

With the introduction of digital tools, various new projects were set up to map the exchange of knowledge and to analyze the creation of scholarly networks in Europe. The beautiful visualizations of the project *Mapping the Republic of Letters* of the Stanford Humanities Research Center made the headlines of the New Yorker. In Europe, Oxford University in the *Cultures of Knowledge* project is building a large repository to make the research on the Republic of Letters available on an international level.

Although we cooperate with these consortia to create a Digital Republic of Letters, one of the two projects discussed here: *Circulation of Knowledge and learned practices in the 17th century Dutch Republic: A Collaboratory around Correspondences* (CKCC)[1] (see also Roorda & Bos & van den Heuvel 2010), is different in geographical range and in its analytical depth.

First CKCC does not follow the correspondences of all European scientists, but of those scholars that lived or sojourned extensively in the Low Countries. The scientific revolution of the 17th century was driven by discoveries at sea, in observatories, in workshops of artisans, and in libraries. The Dutch Republic with its global trade network, its book printing industry, and its relative tolerance to religious differences became a refuge for intellectuals from around Europe: an information society avant la lettre. As such, it is an interesting counterpart to traditional studies in which knowledge production primarily is described as a scientific revolution driven by protagonists in the Galileo-Descartes-Newton tradition.

---

[1] Project website: ckcc.huygens.knaw.nl.

A second difference with the above-mentioned projects is in the depth of analysis. Instead of focusing on metadata to explore the exchanges of knowledge between scholars in Europe and overseas, CKCC focuses on the data, on the letters themselves. It does not only try to answer how knowledge was disseminated in correspondences, but also to establish how new information was picked up, processed, and finally accepted in scholarly communications. What is the impact of the correspondences and how did new scientific topics and scholarly debates around them emerge? To answer those questions CKCC digitized the corporora of published editions and of unpublished letters from the scholars Caspar Barlaeus (1584-1648), Isaac Beeckman (1588-1637), René Descartes (1596-1650), Hugo Grotius (1583-1645), Constantijn Huygens (1596-1687) Christiaan Huygens (1629-1695), Dirck Rembrandtsz van Nierop (1610-1682), Johannes Swammerdam and Anthoni van Leeuwenhoek (1632-1723). Software has been developed to analyze this machine-readable corpus of approximately 20.000 letters to detect topics and to visualize meaningful patterns in the networks of scholars that discuss them by using a combination of text mining, topic modeling, language technology, and visualization techniques. This analysis works in two different ways: a researcher can query the database with specific keywords, which will get a presentation of all the letters in which these words occur. Apart from the fact that all the queried keywords will light up in the text, the computer generates the most frequent words (in a different color) in relation to them. This way a researcher can test hypotheses about the expected outcomes of her/his queries, but at the same time serendipity has a better chance because of the computer generated terms that might convey unexpected meanings, which have to be put into context by additional research.

The complexity of this set is mainly caused by the multiple languages in historical forms that occur in the corpus: Latin, French, Italian, Dutch and English, not to speak about the spelling variants. After several years of experimenting we are entering a phase in which the database can be opened up in a web-based collaboratory and more data added. Moreover, the data is to be enriched with annotations.

We state that the results of these experiments with topic modeling, language detection, and visualization require new knowledge management and archive practices. To that end, we will formulate a new paradigm where annotations will play a key role.

**KNOWLEDGE MANAGEMENT OF MIXED AND PARTIAL DOCUMENTS**
The challenge of CKCC is to study the appropriation of knowledge in an international context and to recognize the development of themes of interest and debates between scholars or in larger networks distributed over space and time. In order to recognize meaningful patterns in the machine-readable corpus, topic modeling is used. It is based on the distribution of words over the text in the documents and is used to find similar words, similar documents or documents similar to arbitrary text. It does this by calculating similarities between words and texts, which constitutes a statistical approach to topics. However, the specific characteristics of the corpus of about 20.000 letters not only complicate the analysis and visualization of meaningful patterns but also require a particular management for pre-processing the datasets for use and re-use in digital humanities research (Roorda & Bos & van den Heuvel, 2010; Wittek & Ravenek 2011).

Letters often address more than one topic and their rhetorical opening and closing phrases are seldom relevant to their content. For that reason it is not only important to be able to segment the letters to the paragraph level but also to exclude certain phrases from content extraction.

Not only are the various digitized corpora so different in format and coding that much data curation is needed to make them suitable for analysis, but also the multilinguality and spelling variations in the letters require additional operations.

The choice of language is very inconsistent over the corpus as a whole. About 95% of all letters are written in Dutch, Latin, and French, the rest in German, Greek, Italian, and English. For some languages, it is a profitable investment to use additional language resources and tools, but not for all of them. Moreover, the letters themselves are not monolingual, but even inside sentences language switches do occur.

These 17th century letters exhibit much spelling variation. For instance, the name of Christiaan Huygens van Zuylichem in the CKCC corpus is spelled in more than 320 different ways. This requires additional language tools and methodologies, such as Named Entity Recognition to improve the recall of queries and of computer generated topics.

In the first phase of CKCC, the topic model of Latent Dirichlet Allocation[2] was used. In this model, documents are considered as random mixtures over latent topics, where each topic is characterized by a distribution of words. Using LDA the computer generated 100 strings of related words; each string was manually labeled by researchers within the team based on their domain expertise.

---

[2]This method and Latent Semantic Analysis and Random Indexing (to be mentioned subsequently) and their application in CKCC are explained in (Wittek & Ravenek, 2012).

After a year of developing the topic model, the database was tested by participants of an international workshop *Mathematical Life in the Dutch Republic*[3]. We asked three test groups of in total circa 20 historians of science to explore the possibilities of this tool and to inquire in what ways it could contribute to their historical research. Although the researchers acknowledged the potential of the database, they came up with two serious problems. They experienced the database as black box which was hard to control, and their queries often had a limited recall. To overcome these problems a mixed strategy was developed for the next phase of the project. Faceted search was improved to enhance the manipulation possibilities of the researchers. Experiments were set up with two different models of topic modeling next to Latent Dirichlet Allocation (LDA): Latent Semantic Analysis (LSA) and Random Indexing (RI) and combined with language normalization. Researchers were involved in labeling the terms (20 in a subset set of 300 letters) that were generated during these experiments to enable an evaluation of the outcomes. It goes beyond the scope of this article to describe these experiments in detail, but best results were achieved with RI in combination with stemming and removal of stop words.

For the implementation, a combination of LSA and RI was used in two scenarios. (1) Query terms of users are forwarded to LSA and RI models that return a ranked list of keywords that are the most relevant to the topic(s) underlying the query. (2) Text fragments specified by users are forwarded to both models and the LSA and RI models return a ranked list of letters that are most relevant to this input. In short, full text search can be enhanced with query terms suggested by the topic model, and it is possible to query for letters that are similar to a given text fragment.

Despite its potential usefulness there is still a long way to go given the multilingual situation and the spelling variants. To improve the recall of keywords, other experiments are set up Involving Named Entity Recognition. Once again, researchers play a role in the evaluation of the automatically generated terms, so that after several iterations of feedback the recall can be improved. Thus, enhancing the queries by topic modeling requires annotation, for the presentation of the results to the end user as well as for the experts' feedback to the software.

**THE NEED TO ANNOTATE AND NEW PARADIGMS IN ARCHIVING NOTES IN HUMANITIES RESEARCH**
Several studies have pointed to the different nature of data in the humanities. They are often multilingual, historically specific, geographically dispersed, and ambiguous in meaning (ACLS, 2006), (Borgman, 2007). Humanities scholars are concerned with the problem of meaning: how it is created, communicated, manipulated, and perceived. In order to contextualize data, they require annotation. Contextualization by annotation has a long history. Famous is the history of the footnote by (Grafton, 1997), but its future is still unclear.

"As the footnote reconfigures itself for the digital world, opportunity and danger are waiting side by side for it" (Zerby, 2003: 144).

Bader stated in *The New York Times*: "Forget Footnotes. Hyperlink. Old Media, Meet New Media". She claimed that after the eviction of the footnote by book publishers, they would find a new home in the hyper-link construction of the World Wide Web. "Indeed the Web has not only revived the footnote, it has spawned a cross-referencing craze that renders the formerly complete media event into a […] wallflower waiting to be courted by the next available annotator" (Bader, 2001).

The statements of Zerby and Bader reveal two problems. We do not know anymore what the function is of the footnote in digital environments, which played such an important role in the contextualization of (humanities) research. Secondly, we are in need of new paradigms to preserve the digital counterpart of the footnote, the annotation by man or by machine, for re-use by researchers. The new role of the footnote in the virtual research environment has hardly been explored. An interesting exception is the study that presented the Multimedia Digital Annotation System, *MADCOW* (Bottoni & al., 2004), in which a functional taxonomy of "content" annotations (explanation, comment, question, example etc.) as opposed to metadata annotations was formulated based on Rhetorical Structure Theory (RST). Such a functional taxonomy can be developed to assign attributes for the contextualization of topics.

Moreover, the MADCOW project signaled the problem that users are limited by either navigating through specific browsers with annotation facilities on a restricted set of contents or have to disrupt their navigation to start an annotation application. The MADCOW tool allowed users to switch between navigating and annotating modalities with the Web content. Here we try to extend these modalities beyond Web content, to include in principal all sorts of documents, and to explore its implications for preservation practices.

**ANNOTATIONS AND DIGITAL SCHOLARSHIP**
The practice of annotating is a traditional ingredient of research. How can annotating support modern, digital forms of research? Is the digital version of an annotation versatile enough to express new results? How do digital annotations behave in the total workflow of exploration, hypothesizing, testing, publishing, and archiving?

In order to gain practical insights in these matters we have considered two significant projects that truly are representative of digital humanities research of which the first one is CKCC, described above, and the second one is rather a programme than a project: *Data and Tradition. The Hebrew Bible as a linguistic corpus and as a literary composition (DTHB)*[4]. This work builds on a multi-decade effort to linguistically markup the complete text of the Hebrew Bible. The result is a text database where morphemes, words, phrases, and higher-level text objects carry many features. A version of this database has been deposited into the DANS archive, where it is stored as a compressed SQL dump (Talstra, 2012). This act happened during the workshop: *Biblical Scholarship and Humanities Computing: Data Types, Text, Language and Interpretation*[5] where an international group of experts reflected on how to bring these resources to better fruition in the digital age. Live versions of this database run on researcher's computers, where they can craft queries of which the results may or may not support specific interpretations of the text. If a linguistic peculiarity shows up in a difficult passage, one can query the database and see whether it is a true exception to the known rules, or just an instance of a regular but rare pattern, to name a typical use case. Hundreds of queries have been crafted, run, and studied, all in relation to interpretation issues.

---

[3] 2010, December 6-10, Lorentz Center, Leiden, tinyurl.com/lorentz-mat-life.

Both CKCC and DTHB have produced curated sources plus analytical results. Yet it is far from clear how these results can partake in a process of accumulation and sharing. Here lies our motivation to explore the power of annotations.

The central statement of this part is that annotations are indeed a powerful carrier of digital scholarship and that they can bridge the gap between past and future research, provided they conform to a generic model that supports preservation and sharing.

In order to substantiate this statement, we have to argue that:

- there are frameworks for web-based, digital annotations;
- annotations are versatile: they can express queries, features, keyword and topic assignments;
- annotations can be made portable: they still make sense when their targets move or change;
- annotations must and can be managed with their metadata, provenance, and types;
- annotations can "drive" end user applications.

Of course, we cannot rigorously prove these assertions. We will draw on our own experiences in building (demo) applications that are driven by queries and features *as annotations* in the DTHB case, and by topics and keywords *as annotations* in the CKCC case.

**Open Annotation Collaboration**

The realization that annotations are important carriers of scholarship, and the fact that in practice annotations tend to become locked up in the systems used to create them, has led to several attempts to standardize annotations and turn them into *web resources*. Two of those attempts, the Annotation Ontology (Ciccarese, 2011)[6] and the Open Annotation Model (henceforth OAM) (Sanderson & van de Sompel, 2011)[7] are currently under consideration of the W3C Open Annotation Community Group[8] with the aim of reconciling the two into a common, RDF[9]-based specification. The guidelines in (Sanderson & van de Sompel, 2011) are particularly concise and revealing. To summarize even more: the OAM focuses on the basic structure of an annotation: a body is taken to comment on one or more targets, and the annotation binds them together. Annotations, body, and targets are all addressable as web resources. They can all have separate metadata, including authorship, but the metadata is not part of the model. The model is agnostic to the specific protocols, platforms, and applications with one exception: everything is geared to the architecture of the web with its HTTP[10] protocol. The implicit consequence is that OAM-annotations can be expressed as RDF and become part of the Semantic Web.

So far, the guidelines reveal that very important goals are being achieved: annotations can be shared easily across applications, platforms, and institutions. They can be discovered, filtered by the metadata they are linked to, and organized by the resources they target, and moved around and aggregated by discovery services.

Yet, the guidelines also point to challenges: (1) real annotations need to target fragments of resources, but how can they be specified in interoperable ways? (2) Resources tend to move and change, so how are the annotations that link to them, either by body or by target, to be maintained? (3) The basic model is bare, and lots of information about annotations has to be expressed in ways not prescribed by OAM, so how much interoperability can be actually achieved?

---

[4] Project *Data and Tradition. The Hebrew Bible as a linguistic corpus and as a literary composition*. Initiated by Eep Talstra, from 2010-07-01 to 2014-06-30. See tinyurl.com/nwo-nl-dthb. More projects in the same programme are listed at tinyurl.com/nwo-nl-talstra.

[5] 2012, February 6-10, Lorentz Center, Leiden, tinyurl.com/lorentz-hum-comp.

[6] See: tinyurl.com/annot-ont.

[7] See: openannotation.org.

[8] See: tinyurl.com/w3-annot.

[9] RDF: Resource Description Framework. The language of the Semantic Web, also known as Linked Data. See linkeddata.org.

[10] HTTP: HyperText Transfer Protocol. Defined here: tinyurl.com/ietf-http.

From the perspective of a research archive, which preserves resources past their active lifetime in an encapsulated form, in order to revive them when somebody is interested in them, exactly these two issues of addressing and metadata are of utmost importance.

In our view (1)+(2) are fundamental issues that require additional concepts. We address them in section *Portable Annotations*. As to (3), there is a general tendency in archives, repositories, and cultural heritage institutions to conform their metadata to the ontologies that are being designed on the Semantic Web, not only for the metadata profiles, but also for the actual values that metadata fields may take (Gradmann, 2010). OAM is very well poised to take advantage of these developments, since it is itself defined in Semantic Web terms.

**Queries, Features, Topics and Keywords as Annotations**

As discussed above, the results of CKCC and DTHB are predominantly queries and features (DTHB) and topics and keywords (CKCC). Here we explain how we translated these items all into annotations. We subsequently wrote two web applications that present these annotations next to the resources in one interface.

QFA (Queries/Features as Annotations)[11] (figure 2) is written for DTHB, and TKA (Topics/Keywords as Annotations)[12] (figure 3) is written for CKCC material.

The intention was to explore if one could build usable interfaces that are driven by annotations, and with limited effort. To this end we developed two end-user applications that directly operate on sets of annotations using the abstract model, and connect them with data sources that they are about. We assume that both data sources and annotations have been previously imported into relational databases. (See further, *Portable Annotations* below).

*Queries as Annotations*
Queries are active, dynamic forays into landscapes of data. Annotations are passive, static comments on fragments of data. What do they have in common?

One might expect that we are preserving queries with the aim to be able to run the query over and over again for the indefinite future. Or do we? It would require that we remain familiar with that version of the query language, and with the corresponding version of the database system forever and ever. It will become increasingly difficult to compare those query results with later ones because the modern query will not run on the old system and vice versa. The matter is not academic. In this particular case, the queries are expressed in MQL, which is an implemented version of QL, defined by (Doedens, 1994) as a query language specifically geared to text databases[13].

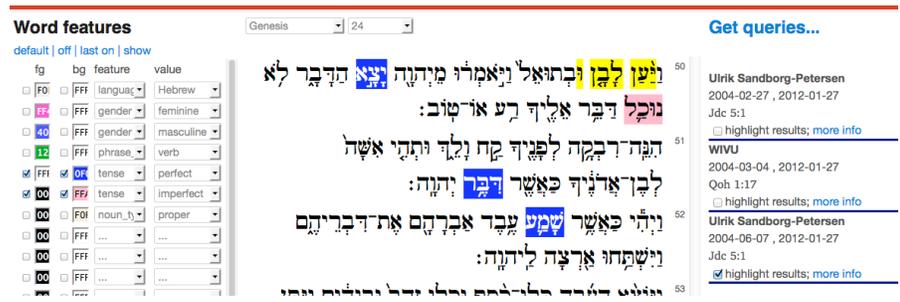

**Figure 1. Screenshot of Queries/Features as Applications**

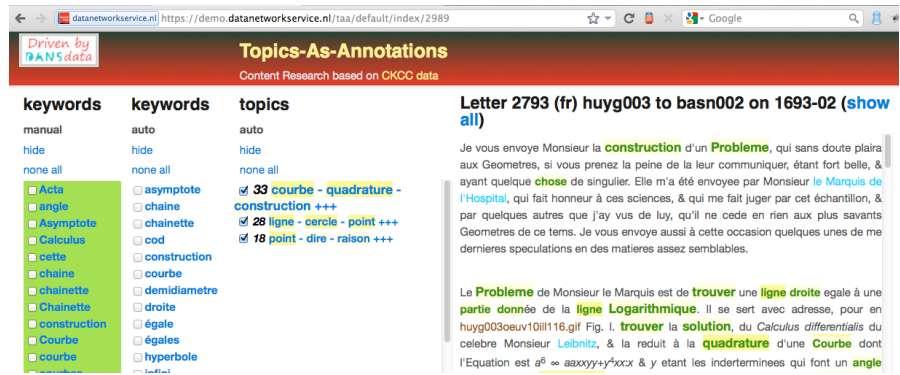

**Figure 2. Screenshot of Topics/Keywords as Annotations**

Although the implementation, EMDROS (Petersen, 2004)[14] is open source, well documented, and a powerful solution for text databases, it is definitely not a mainstream application, and its life span is hard to guess.

For preserving the results of scholarship, there is a better option. We can select the important queries, those that have been used to obtain new interpretations that have been published in journals. The query instruction is then the *body* of an annotation, and the query results are the (many) *targets* of that same annotation. Annotations will be linked to metadata specifying the related research problem, the author of the query, and the moment of its last run. That will give the future user a good picture of past research. In addition, in current research users can stumble upon query results as targets of annotations, so that these annotations lead them from passages to queries, exactly in the opposite direction that one usually follows with queries. It is the direction of serendipity.

---

[11]Application: tinyurl.com/demo-taa, wiki: tinyurl.com/wiki-taa.

[12]Application: tinyurl.com/demo-qaa, wiki: tinyurl.com/wiki-qaa.

[13]The acronym QL may best be read as: QUEST-like Query Language, and MQL stands for Mini QL. Appendix 1 of (Doedens, 1994) contains a historical account.

[14]See emdros.org.

*Features as Annotations*
In the DTHB case features are linguistic properties of the form *key=value* that apply to text objects of nearly every granularity, from morpheme through part-of-speech up to book. These features are the product of many years of manual labor, combined with automatic processing. They have been checked and revised. They constitute a treasure trove. They live in the same implementation of text databases, Emdros, as the queries above. By transforming the features into annotations, we potentially unlock the value that is hidden here.

In this case, we simply chose as bodies strings of the form *key=value*. The targets are the objects that carry that feature value.

In our demo application QFA we give the user 70 *key=value* combinations at the word level to play with. As an example, a user can tell the application to show all verbs with *tense=imperfect* in blue and all verbs with *tense=perfect* in red. This helps to interpret narrative structures, even if you do not know Hebrew, although being a linguist helps.

Again, this is a case of annotations with (very) many bodies: the annotation with body *gender=masculine* has 101335 targets! The number of targets of *gender=feminine* is left as an exercise to the reader.

*Topics and Keywords as Annotations*
Extracting topics from texts is as useful as it is challenging. Topics are semantic entities that may not have easily identifiable surface forms, so it is impossible to detect them by straightforward search. Topics live at an abstraction level that does not care about language differences, let alone spelling variations. Therefore, if one has a corpus with thousands of letters in several historical languages, and wants to know what they are about without actually reading them all, a good topic assignment is a very valuable resource indeed.

There are several ways to tackle the problem of topic detection, and they vary in the quality of what is detected, the cost of detection, and the ratio between manual work an automatic work. Several of these methods have been (and are being) tested CKCC as explained before.

It is not the purpose of this paper to go into topic modeling in depth. Here we are concerned with gathering results, even intermediate results, and making them re-usable for subsequent attempts to uncover the semantic contents of the corpora involved. For our demo application, we gathered three kinds of intermediate results: (1) automatic keyword assignments, (2) manual keyword assignments, (3) automatic topic assignments detected by a specific algorithm. We used the complete corpus of letters from and to the Dutch 17$^{th}$ century scholar Christiaan Huygens (3090 letters).

The mapping from keyword assignments to annotations is simple: bodies are the keywords; targets are the letters to which the keywords are assigned. There is no fragment addressing here.

Topics reveal two complications when translating them into annotations. (1) A topic is not a single word but a complex object in itself. In this context, it is a collection of words that span a semantic field. Moreover, each word contributing to a topic does so with a certain relative weight. (2) When a topic is assigned to a letter, the assignment has a certain confidence, expressed as a number. This could be modeled as an extra annotation on top of the annotation that merely links a topic to a letter: the extra annotation has the confidence as body and the other annotation as target. In our application, however, we have opted to include topic and confidence in one body, as distinct fields. There are even more options, for which we refer to the wiki about TKA[15].

**Portable Annotations**

*Beyond RDF*
So our annotations are not coded in RDF, they have no URIs[16], and they do not conform to the Linked-Data aspects of OAM. There are good reasons for this: neither the sources nor the annotations that result from CKCC and DTHB are currently web resources.

Nevertheless, there is a sense in which we conform to OAM: the annotations reside in a different database than the sources do, and the link between annotations and their targets is strictly symbolic, not dependent on database modeling and technology (no foreign key constraints).

One could say that we enforce *modularity* between sources and annotations, in the sense that annotations can be ported from one source to a comparable source. From here it is not a big step to completely conform to OAM: (1) import real RDF annotations to local database tables from where they drive local applications; (2) if a local application produces annotations that must be shared: export them as RDF. In both cases, local addresses must be translated into absolute URIs.

*Usefulness of Porting Annotations*
Now we arrive at a tempting picture: annotations that are *portable*. Many sources are available in several versions, in many copies, in different formats, in multiple languages, and in diverse media. Many annotations on a resource still make sense if one explores other variants of it. Here are some examples:

(1) (from DTHB): there are various authoritative versions of the Hebrew text. We have compared the Biblia Hebraica Stuttgartensia (BHS)[17] with the Westminster Leningrad Codex (WLC)[18]. Most of the differences are different word divisions and different diacritical marks. That means that the vast majority of Feature and Query annotations based on the BHS also apply to the WLC. Moreover, there is a set of features, by a different enterprise (Groves & Lowery 2006)[19], based on the WLC, which can also be applied to the BHS. Even the mismatches are interesting!

(2) (from DTHB): there are word-by-word translations of the Hebrew Bible into English. For non-Hebrew-readers, it might be interesting to see which words in such a translation derive from a masculine and which from a feminine word. Such an observation can be easily achieved if we can port the feature annotations from the Hebrew source to such a translation.

(3) (from CKCC): the manual topic assignments are a valuable resource. New attempts at topic modeling could make good use of that, for training or testing purposes. In those cases, it would be convenient to retrieve such annotations from an archive and then to be able to reapply them on new incarnations of the sources.

---

[15] tinyurl.com/wiki-taa-topics.

[16]URI: Uniform Resource Identifier, which can be *dereferenced* by means of the HTTP protocol. The definition URI is at tinyurl.com/ietf-uri.

| concept | example | characteristic |
|---|---|---|
| work | Beethoven's Ninth Symphony | distinct creation |
| expression | musical score | specific form |
| manifestation | recording by the London Philharmonic in 1996 | physical embodiment |
| item | record disk | concrete entity |

Table 1. FRBR's view of the world

*URIs, Anchors, FRBR*
OAM requires that annotations point to their bodies and targets in the Linked Data way: by proper HTTP URIs. If the resources in question are stable and being maintained by libraries, archives and cultural heritage institutions, it becomes possible to harvest many sorts of annotations around the same sources. This is an organizing principle that is quite new and from which huge benefits for data mining and visualization are to be expected.

[17] tinyurl.com/bhs-browse.

[18] tinyurl.com/tanach-tech.

[19] The Westminster Hebrew Morphology. tinyurl.com/groves-whm.

In practice, however, there are several scenarios in which (fragments) of resources are not addressed in a stable way. This happens for instance when resources go off-line into an archive. In case we want to restore those resources later on, the means of addressing them from the outside may have changed. Moreover, there might not be a unique, canonical restored incarnation of that resource. For that reason one needs *anchors* to resources that enable the re-use of annotations that have been archived in the past.

The solution adopted in QFA and in TKA is to work with *localized* addresses. These are essentially relative addresses that point to (fragments) of local resources that are part of a local corpus.

There is an ontological consideration involved here. The model of *Functional Requirements for Bibliographic Records* (IFLA, 1997-2009) makes a distinction between work, expression, manifestation, and item. *Work* is a distinct intellectual or artistic creation. As such, it is a non-physical entity. *Expression*, *manifestation,* and *item* point to increasing levels of concreteness, an item being a concrete entity in the physical world. Wikipedia[20] gives a nice example from music, see table 1.

The full refinement of these four FRBR concepts is probably not needed for our purposes. Yet, a distinction between the *work*, which exists in an ideal, conceptual domain, and its *incarnations*, which exist in physical reality, is too important to ignore. It bears on the ways by which identifiers to works and incarnations can be kept stable. Identifiers to works identify within conceptual domains, but they have no function in physically locating works. These identifiers are naturally free of those factors that make a typical hyperlink such a flaky thing. So whenever annotations are about aspects of a resource that are at the *work* level, they have better target those resources by means of *work* identifiers. Moreover, the distinction between work and incarnation also applies to fragments of works. Most subdivisions, such as volumes, chapters, and verses in resources do exist at the work level, albeit that there are some fragments that are typically products of the incarnation level, e.g. lines and pages.

We can now define our anchors as identifiers at the work level, for resources and their fragments. This is in fact the nature of our *localized addresses*.

Quite often, the sources themselves and their fragments have anchors that are recognized by whoever is involved with them. Take the books, chapters, and verses in the bible, for example. Even where there are no universally recognized anchors, it is easier to translate between rival anchoring schemes, than to maintain and multiply stable identifiers at the incarnation level.

[20] tinyurl.com/wikip-frbr.

Lurking below the surface there is the question: to what extent are differing versions incarnations of the same work? Can we keep fragment identifiers stable under versioning? This is really a complex issue, and we plan to devote a completely new demo application to it in a new use case. (See the wiki on Portable Annotations)[21].

*Statement*
Not all variance between sources can be productively addressed with time-based versioning. There are deeper reasons for variation and deeper reasons for identification than sequences of surface forms.

If we ignore those reasons, and if we omit to base our identifiers on them, we will not have truly portable annotations.

**Annotation management: metadata, provenance and types**
The role of metadata for annotations is (at least) twofold: first, they enable to assess the quality, significance, and meaning of an annotation. Quality judgments can be made based on the provenance: who made the annotation, for which project, when? Significance can be gleaned from a list of publications that are associated with that (set of) annotations. Meaning can be retrieved from pointers to reference materials.

As OAM is firmly integrated in the Semantic Web effort, there are no conceptual limitations on linking metadata to annotations.

The second role of metadata is to enable annotation-driven applications to decide how to best filter and display the annotations. Here the typology of annotations comes in. We exposed four not too ordinary types of annotation, each with its own requirements for display. The unlimited linking of metadata to annotations is problematic for generic applications. How do applications recognize what metadata is available and by which metadata they should let themselves be controlled?

Here we find ourselves on the middle ground between the rigor of what is within the limits of OAM and the polymorphism of what lies outside it. For dedicated applications, there is no problem: you can tell them where to look, but fully generic annotation-driven applications will have difficulties here.

**Annotation-driven applications**
How difficult is it to develop an annotation-driven application that deals with significant amounts of data and annotations, and that presents a usable interface to the end user?

---

[21] tinyurl.com/wiki-pa. Beware that this is work in progress.

*Design*
The demo applications QFA and TKA are driven by a database containing the source materials and a separate database with the annotations. There is no mingling or tight coupling between the sources and the annotations. The only links are the anchors: symbolic expressions in the annotation targets that refer to fragments of the sources.

*Functionality*
Both applications display the source material in a broad column, and the annotations in narrower columns next to the sources. The targets of the annotations can be highlighted in the sources, and the user has some control over the highlighting, depending on the type of annotation.

We invite the reader to explore these applications to get a more detailed picture.

In short, these applications visualize the annotations and the sources in basic, not too crude ways, adapted to the different kinds of annotation.

*Implementation*
In order to rapidly implement our ideas concerning annotations and sources we needed a simple but effective framework on which we could build data-driven web applications. We found it in the shape of Web2py (Di Pierro, 2007-2010).

We needed very little code on top of the framework, just a few hundred lines of Python and Javascript each. Deployment of these apps is completely web-based, and only takes seconds.

Most work went into the data preparation stage, where we used Perl and shell scripts to compile data from various origins into SQL-imports for sources and annotations. These scripts were also in the few hundred lines range.

*Missing Link*
What these demos still lack is full RDF capability. Once these sources are truly web resources, we expect that it is easy to make an import/export facility to turn database annotations into real RDF annotations. How to translate our fragment anchors into HTTP URIs is still an open question. Finally, work is to be done in order to get the best of the worlds of relational databases and of linked data, see e.g. (Baron & Di Pierro, 2010).

**CONCLUSION**
We have investigated the feasibility of using annotations as portable carriers for diverse results of scholarship in the humanities. We found that annotations are versatile enough to carry the products of digital scholarship such as query results, features, topics, and keywords. The Open Annotation Model represents annotations as web resources, which makes them easy to share beyond the systems in which they originated. Annotations can be managed by unlimited association of metadata. The development of annotation-driven applications is doable: the focus remains on the data, and does not shift to the software.

Yet, the web-based model for annotations is not fully compatible with the process of archiving and re-use. This would greatly be improved if we could make annotations more portable across variant resources. And that, in turn, boils down to using anchors for targeting resources and their fragments. Anchors are identifiers at the work-level in the FRBR sense.

Let us briefly consider what this outcome means for digital humanities in general.

In the non-digital ages before us, scholars relied on harmonization efforts such as standard editions of historical texts, because the source materials were simply too complex to deal with in their raw form. It had the character of projecting the data on a space of one dimension. Now there is a growing pressure to investigate (again) the raw data, find new perspectives, and preserve the connections between interpretations and data in a much more transparent way. This shift in research paradigm can only succeed if it is matched by a shift in archiving methods.

Annotations have the potential to unlock data that is behind the barriers of application interfaces and data models. They facilitate deep linking to fragments. They can be instrumental in identifying interesting slices of the data that could not be accessed as such before. This is particularly useful in disciplines whose business it is to make distinctions between objective data and many layers of interpretation, where those interpretations are based on the data themselves in combination with any amount of data from the context.

The fabric of objects and meanings that humanities research is creating must be taken care of in such a way that it remains navigable from all imaginable entry points in all conceivable directions. We have shown that annotations are up to the task. Their way into the web of linked open data is being paved. If, in that process, they can play nice with the distinction between concept and realization, they constitute a new archiving paradigm.

**ACKNOWLEDGMENTS**
Walter Ravenek (Huygens ING) for helpful comments on topic modeling; Eko Indarto (DANS) for helping to develop a first version of QFA in very short time; Andrea Scharnhorst (DANS) for granting additional time for research; Joris van Zundert (Huygens ING) for facilitating an inspiring Interedition bootcamp[22] which set me (Dirk) on the track of rapid development.

---

[22] tinyurl.com/intered-lvn.

**REFERENCES**

ACLS (2006). Our Cultural Commonwealth: The Report of the American Council of Learned Societies' Commission on Cyberinfrastructure for Humanities and Social Sciences. Retrieved 2012-04-28 from http://www.acls.org/cyberinfrastructure/OurCulturalCommonwealth.pdf

Bader, J.L. (2000). Forget Footnotes. Hyperlink. The New York Times, Sunday 16 July 2000 Section 4 Week In Review.

Baron, C., Di Pierro, M. (2010). Publishing Linked Data Using web2py. School of Computing, DePaul University of Chicago. Retrieved 2012-04-28 from tinyurl.com/web2py-ld-article (pdf).

Bottoni, P., Civica, R., Levialdi, S., Orso, L., Panizzi, E., Trinchese, R. (2004). MADCOW: a Multimedia Digital Annotation System. In M.F. Costabile (Ed.), *Proc. Working Conference on Advanced Visual Interfaces (AVI 2004)* (pp. 55-62). New York: ACM Press.

Borgman, C. (2007). Scholarship in the Digital Age. Information, Infrastructure and the Internet, Cambridge (Mass.), London: The MIT Press.

Ciccarese, P., Ocana, M., Castro, L.J.G., Das, S., Clark, T. (2011). An Open Annotation Ontology for Science on Web 3.0. J. Biomed Semantics 2011, 2(Suppl 2):S4 (17 May 2011).

Di Pierro, M. (2007-2011). web2py. Full Stack Web Framework, 4th edition. Online book. Retrieved 2012-04-28 from web2py.com.

Doedens, C.F.J. (1994). Text Databases. One Database Model and Several Retrieval Languages. *Language and Computers, Number 14*. Editions Rodopi Amsterdam. Amsterdam and Atlanta, GA. ISBN: 90-5183-729-1.

Gradmann, S. (2010). Knowledge = Information in Context: on the Importance of Semantic Contextualisation in Europeana. White paper. Retrieved 2012-04-28 from tinyurl.com/europeana-gradmann (pdf).

Grafton, A. (1997). The Footnote. A curious history. Cambridge (Mass.): Havard University Press.

Groves, A., Lowery, K., (Eds). (2006). The Westminster Hebrew Bible Morphology Database. Philadelphia: Westminster Hebrew Institute.

IFLA (International Federation of Library Associations and Institutions) (1997-2009). Functional Requirements for Bibliographic Records. Final Report. Retrieved 2012-04-28 from tinyurl.com/ifla-frbr (pdf).

Petersen, U. (2004). Emdros - a text database engine for analyzed or annotated text. *Proceedings of COLING 2004*. 1190–1193. Retrieved 2012-04-28 from tinyurl.com/emdros-coling (pdf).

Roorda, D., Bos, E-J., van den Heuvel, C. (2010). Letters, Ideas and Information Technology. Using digital corpora of letters to disclose the circulation of knowledge in the 17th century". In *Digital Humanities Conference Abstracts King's College London 7-10 July 2010* (pp. 211-214).

Sanderson, R., van de Sompel, H. (Eds.). (2011). Open Annotation: Beta Data Model Guide. Web document. Retrieved 2012-04-28 from openannotation.org.

Talstra, E., Sikkel, C., Glanz, O., Oosting, R., Dyk, J.W. (2012). Text Database of the Hebrew Bible. Dataset available from Data Archiving and Networked services after permission of the depositor through Retrieved 2012-04-28 from tinyurl.com/dans-wivu.

Wittek, P., Ravenek, W. (2011). Supporting the exploration of a corpus of 17th century scholarly correspondences by topic modeling. In B. Maegaard (Ed.), *Proceedings of Supporting Digital Humanities 2011: Answering the unaskable*. Copenhagen.

Zerby, C. (2003). The Devil's Details: A History of Footnotes. New York: Touchstone.